\documentstyle[aps,preprint,prb,psfig]{revtex}
\newcommand{\RR}{\rangle \rangle}
\newcommand{\LL}{\langle \langle}
\newcommand{\RA}{\rangle}
\newcommand{\LA}{\langle}
\begin{document}
\title
{Spin Models on Non-Euclidean Hyperlattices:  Griffiths Phases
without Extrinsic Disorder}
\author{J.C. Angl\`es d'Auriac,$^1$ R. M\'elin,$^1$ P. Chandra$^2$
and B. Dou\c{c}ot$^3$}
\address{$^1$CRTBT-CNRS, 38042 Grenoble Cedex, FRANCE}
\address{$^2$NEC Research Institute, 4 Independence Way, Princeton NJ
08540, USA}
\address{$^3$ LPTHE-CNRS UMR 7589, Universit\'es Paris 6 et 7,
4 Place Jussieu, 75252 Paris Cedex 05, FRANCE}
\maketitle
\begin{abstract}
We study short-range ferromagnetic models residing on planar manifolds
with global negative curvature.  We show that the local metric
properties of the embedding surface induce droplet formation
from the boundary, resulting in the stability of a Griffiths phase
at a temperature lower than that of the bulk transition.
We propose that this behavior is independent of order
parameter and hyperlattice specifics, 
and thus is universal for such non-Euclidean spin models.
Their temperature-curvature phase diagrams 
are characterized by two distinct bulk
and boundary transitions; each has mean-field critical behavior
and a finite correlation length related to the curvature of
the embedding surface.
The implications for experiments on superconducting hyperlattice
networks are also discussed.
\end{abstract}

\pagebreak

\section{Introduction}

Disordered spin systems behave in ways qualitatively
distinct from their periodic
counterparts.\cite{Anderson79,Ramakrishnan87,Mydosh93}  
More specifically,
they display broad relaxation spectra,
in clear contrast to the
Debye relaxation observed for spin crystalline materials.  Relaxation
times depend on local conditions, and thus it is
not surprising that randomly-coupled spins relax
on a wide distribution of time-scales.  There are 
now several anomalous magnetic materials\cite{Aeppli97,Ramirez99,Wills00} 
whose non-Debye relaxation
seems
to be 
determined by lattice topology\cite{Chandra93} rather than by disorder.  The
theoretical challenge is to identify and characterize periodic
reductionist models that exhibit such behavior in the absence of
intrinsic randomness.

There are several analytic studies of periodic glass
models,\cite{Bouchaud98}
though their infinite-range nature makes their
relevance to real materials unclear.  More specifically,
the absence of any length-scale in these approaches implies
relaxation on just a few time-scales.  
Detailed analytic methods to treat short-range glasses, even
with intrinsic disorder, remain to be found.  It has
been proposed that their slow dynamics result from rare,
locally-ordered spatial regions that are energetically
probable due to random spin-spin couplings,\cite{Bray86,Fisher86}
similar to the situation associated with Griffiths
phases\cite{Griffiths69}
in diluted ferromagnets;
here large fluctuating droplets lead to relaxation on long time-scales.
Though the application
of this droplet scenario to experimental spin glasses remains 
controversial,\cite{Weissman93}
it is an appealing starting point for geometrically-induced
glassiness.
In quenched ferromagnets, the random initial configuration approaches
equilibrium via domain wall motion.\cite{Bray94} 
This coarsening occurs slower
when the boundaries are pinned by impurities, as is the case in
the random-field Ising model.\cite{Alberici-Kious98}  
Thus one can ask whether local geometry
can induce such slow domain wall dynamics for a short-range
ferromagnetic model. Self-similar lattices are excellent candidates,
since all minority droplets will be non-compact;\cite{Kirkpatrick79} 
indeed non-trivial
slow relaxation has been reported in this case.\cite{Melin96,Butaud98}  
It is therefore
natural to continue this program by studying ferromagnets on lattices
embedded in surfaces with constant negative curvature.  
Dynamics
on such hyperbolic manifolds, particularly in the area of 
chaos,\cite{Balazs86,Gutzwiller90}
are known to be qualitatively similar to those observed in disordered
systems\cite{Comtet00} and thus such surfaces provide promising
settings for the identification of
broad relaxation spectra in disorder-free
models. 

In this paper we show that periodic ferromagnetic models
on hyperlattices with open boundary conditions exhibit
slow dynamics on a distribution of time-scales in the absence
of intrinsic randomness.  Their metric structure makes
it energetically probable for large domains of minority
spins to nucleate at the boundaries, leading to the stability
of a Griffiths phase at a temperature lower than the bulk
temperature.  In dilute ferromagnets with correlated disorder,
the presence of such
rare droplets leads to a diverging magnetic
susceptibility,\cite{McCoy69}
a phenomenon that we also observe for the periodic hyperlattice
models.
We attribute this behavior to the distinction between
bulk and boundary sites, where the latter comprise a significant
proportion of the total site number.  Two mean-field transitions
are identified, associated with the bulk and the boundary
respectively, whose characters are determined by local metric
properties; for example the correlation length associated with
each transition is related to the curvature of the embedding
surface.   We conjecture that this phase behavior is universal
for {\sl all} short-range ferromagnetic models residing on
lattices embedded on hyperbolic planes, independent of details
associated with the spin order parameter or lattice specifications.

The outline of this paper is as follows.
In the next Section (Section II) we discuss geometrical properties of
lattices embedded in surfaces of negative curvature. The model
and its specific heat are presented in Section III. 
The Bethe-Peierls transition for the central spin, that
which is deepest in the hyperlattice, is derived in Section IV.
This analysis is generalized in Section V to an arbitrary lattice
site.  Analysis of magnetization distributions is presented
in Section VI, with particular emphasis on the inferred slow dynamics.
In Section VII, we summarize our results with
a conjectured phase diagram for these
models in the curvature-temperature plane and discuss possible
experiments on Josephson junction arrays.  In order to maintain
the flow of the text, we have delegated the derivation
of the main analytical results to Appendices.

\section{Hyperlattices:  Regular Lattices on a Hyperboloid}
\label{sec:geometry}

\begin{figure}
\centerline{\psfig{file=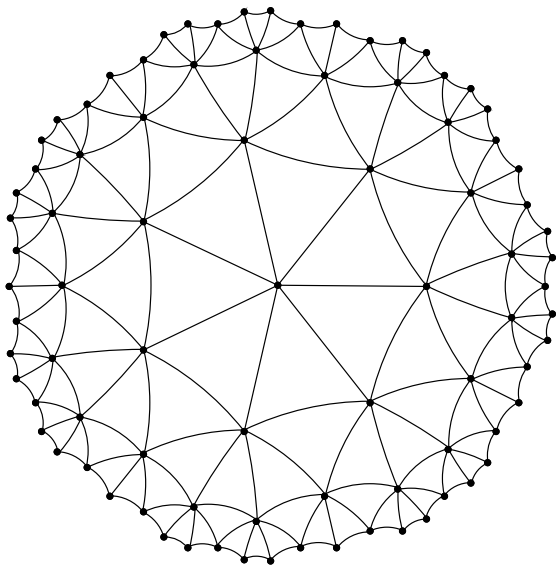,height=8cm}}
\caption{The site centered (3-7) hyperlattice with 3 generations.}
\label{fig:37}
\end{figure}

Infinite regular lattices are characterized by two
integers $(p,q)$ where $p$ is the number of
polygon edges  and $q$ is the
number of polygons around a given 
vertex. On an Euclidean plane (i.e. no curvature), there
are three tilings  corresponding to $(p-2)(q-2)=4$:
the square $(4,4)$,
the triangular $(3,6)$, and the
the honeycomb $(6,3)$  lattices.
On the sphere ${\cal S}_2$  (positive curvature)
the 
five platonic polyhedra correspond to the condition
$(p-2)(q-2)<4$:
the tetrahedron (3,3), the cube (4,3), the octahedron (3,4),
the dodecahedron (5,3), and the icosahedron (3,5).
Hyperlattices, tilings of the hyperbolic plane with
negative
curvature, correspond to the condition $(p-2)(q-2) > 4$;
clearly there exist an infinite number of possibilities.
The site centered $(3,7)$ hyperlattice is shown on
Fig.~\ref{fig:37}.

Loopless trees, such as Bethe lattices,
correspond to regular tilings of the hyperbolic plane
labeled by the integers $(\infty,q)$.\cite{Clark80,Mosseri82}
Hyperlattices can be built in a shell
structure, where each layer is analogous
to a generation of a tree.  For a hyperlattice
of $n$ shells, the number of boundary sites, $\delta V_n$,
scales with the number of bulk sites, $V_n$; since
$\delta V_n \sim V_n^{1-\frac{1}{d}}$ for a Euclidean
tiling, the hyperlattices are considered to have $d=\infty$.
However they possess intrinsic length scales determined
by their radii of curvature, a feature that 
influences the critical behavior of models residing on 
these surfaces.\cite{Rietman92}

It is useful to study the continuous-space
analogue of the hyperlattice, represented
as a unit disc on the Lobachevskii plane
with a metric
\begin{equation}
\label{eq:metrique}
g(z) = \frac{1}{\left( 1 - |z|^2 \right)^2}
,
\end{equation}
where $z=x+ i y$ is a point on the disc.
A finite-size system is obtained
by restricting $z$ to the disc $|z|<R<1$.
The minimum-length paths, e.g. the geodesics,
between two points at coordinates
$z_1$ and $z_2$ are circles perpendicular to
the unit circle.  The distance 
between $z_1$ and $z_2$ is
\begin{equation}
\label{eq:dist}
d(z_1,z_2) = \tanh^{-1}{\left| \frac{z_1 - z_2}
{1 - z_1 \overline{z}_2} \right|}
.
\end{equation}

In order to illustrate the phenomenology associated with different
curvatures, we determine the short-scale corrections
to the perimeter of a circle, ${\cal L}$,  of radius $d$ for the
negative and positive cases. 
First we consider the circle embedded on a two-dimensional sphere
where its perimeter is decreased by the presence of a {\sl positive}
curvature
${\cal R}$:
\begin{equation}
\label{eq:L-R-pos}
{\cal L}_{ {\cal S}^2} = 2 \pi d - \frac{\pi}{3
{\cal R}^2} d^3
+ ...
.
\end{equation}
We can perform an analogous expansion
for the case of the hyperbolic plane with 
the metric Eq.~\ref{eq:metrique}.
The set of points
at a fixed distance $d$ away from a given point $X$ 
(chosen on the real axis) is
found to be represented by the circle
$z=x_0 + R_0 \exp{(i \phi)}$, with
\begin{equation}
\label{eq:x0-R0}
x_0 = X \frac{ 1 - \tanh^2{d}}{1 - X^2 \tanh^2{d}}
\mbox{, and }
R_0 = (1 - X^2) \frac{ \tanh{d}}{ 1 - X^2 \tanh^2{d}}
.
\end{equation}
 The periphery of the circle $z=x_0 + R_0 \exp{(i \phi)}$
is
\begin{equation}
\label{eq:L-R-neg-integ}
{\cal L}_{\rm bulk} = \int_0^{2 \pi} \sqrt{ g(z)} R_0 d \phi
= \pi \sinh {(2d)}
.
\end{equation}
so that a short-scale expansion to order $d^3$ leads to
\begin{equation}
\label{eq:L-R-neg}
{\cal L}_{\rm bulk} = 2 \pi d + \frac{4 \pi}{3} d^3 + ...
\end{equation}
Comparing Eq.~\ref{eq:L-R-neg} with
Eq.~\ref{eq:L-R-pos}, we see that
the curvature associated with the
metric Eq.~\ref{eq:metrique} is ${\cal R} = -1/2$.
On large length-scales, the sphere  
is compact; there are no more points
at a distance larger than a critical value.
By contrast, on the manifold with negative curvature,
the number of
points at a distance $d$ away from a given point
increases exponentially with
$d$ above the curvature radius:
\begin{equation}
\label{eq:L-bulk}
{\cal L}_{\rm bulk} \sim \frac{\pi }{2} \exp{(2 d)}
,
\end{equation}
which is obtained from Eq.~\ref{eq:L-R-neg-integ}.
On a Bethe lattice, the number of points
at a given distance $d$ scales like
$z^d$, with $z$ the branching ratio.
This indicates a close link between the
tree and hyperlattice structures, related
to the underlying manifold with negative curvature.

\section{The Free Energy of the Hyperlattice Model}
\label{sec:simul}

Here we study the nearest-neighbor ferromagnetic Ising model (FIM)
with Hamiltonian 
${\cal H} = - \sum_{\langle i,j \rangle} \sigma_i\sigma_j$,
where $\sigma = \pm 1$ and $\langle i,j \rangle$ are neighboring sites
on a hyperlattice.   A characterization of a special case, the
FIM on the Cayley tree, has already been 
reported.\cite{Melin96,Melin96a,d'Auriac97}
However there exists a temperature scale, $T_g$, below which
large droplets of flipped spins proliferate from the boundaries,
resulting in non-Gaussian magnetization and glassy
behavior for a macroscopic number of 
sites.  In this paper we investigate such boundary-induced
Griffiths phases for periodic
spin models on more general hyperlattices.

\begin{figure}
\centerline{\psfig{file=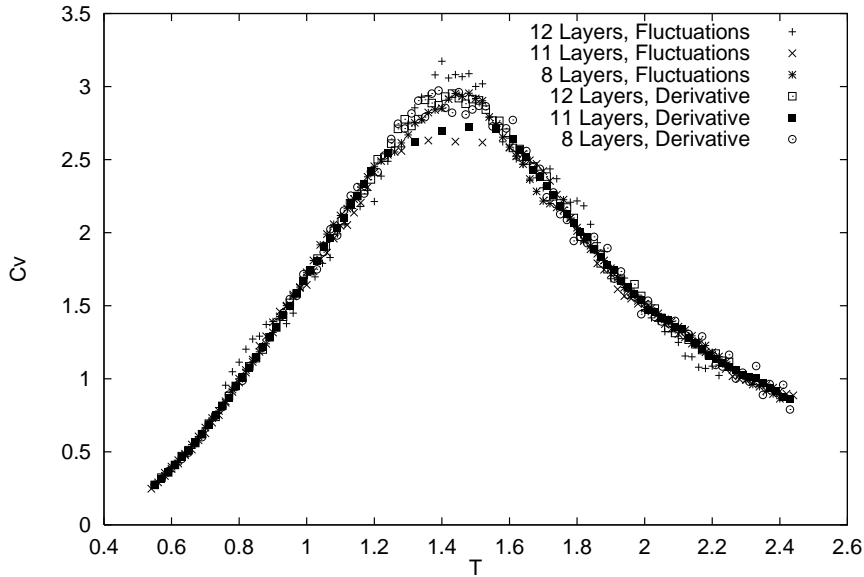,height=8cm}}
\caption{Specific heat for a (3,7) hyperlattice computed
from the numerical
derivative of the internal energy and from the fluctuation of
energy. The agreement indicates that thermal equilibrium
is properly sampled.
The results presented are obtained with
8, 11 and 12 layers.  The number of sites is
4264 (8 generations), 76616 (11 generations),
and 200593 (12 generations). The number of bonds is
10150 (8 generations), 182490 (11 generations),
and 477799 (12 generations).
No sign of divergence is found
when the system size is increased.}
\label{fig:cv}
\end{figure}

The free energy per spin of the ferromagnetic Ising model on the
Cayley tree is analytic for all temperatures; because of the absence
of loops on this pseudo-lattice, it can be obtained from a
high-temperature
series expansion and is
$f(\beta)=-2 \ln{[\cosh(\beta)]}$.
The situation could be different for a general hyperlattice due to
the presence of loops.
The number of independent cycles $n_c$ of a graph with $n_s$ sites and
$n_b$ of bonds is $n_c =  n_b - (n_s-1)$. For
the $(3,7)$ hyperlattice, we have
$n_c / n_s \sim (5-\sqrt 5)/2 \approx 1.38197$, which is even larger than
the square lattice value
$n_c / n_s= \sim $.
Thus the free energy of the FIM on the hyperlattice could
have a singularity, and its absence/presence must be checked
explicitly.
In order to do this, we have measured 
its internal energy $\LA e \RA$ and specific heat
$c_v=(\LA e^2 \RA - \LA e \RA^2)/T^2$ using numerical simulations.
We have ensured that the equilibrium 
distribution was properly sampled
by comparing the numerical
estimate of $d \LA e \RA / d T$ and
$(\LA e^2 \RA-
\LA e \RA^2)/T^2$ and have
used both heat-bath and cluster
algorithms~\cite{Swendsen86}, with
perfect agreement.
The temperature-dependence of the
specific heat is shown on Fig.~\ref{fig:cv} 
for an increasing number of layers $n=8,11,12$. All these curves
display a maximum around the same temperature, but
these peaks do {\sl not} increase
with the system size; thus we interpret this maximum 
as a Schottky anomaly rather as a signature of a phase transition
in the thermodynamic limit.
The free energy of the FIM on the hyperlattice thus 
appears to be analytic at all temperatures, similar to
the situation for the special case of the Cayley tree.
We note that here we are always considering the case of
open boundaries.

\section{The Bethe-Peierls Transition of the Central Spin}
\label{sec:BP}

\begin{figure}
\centerline{\psfig{file=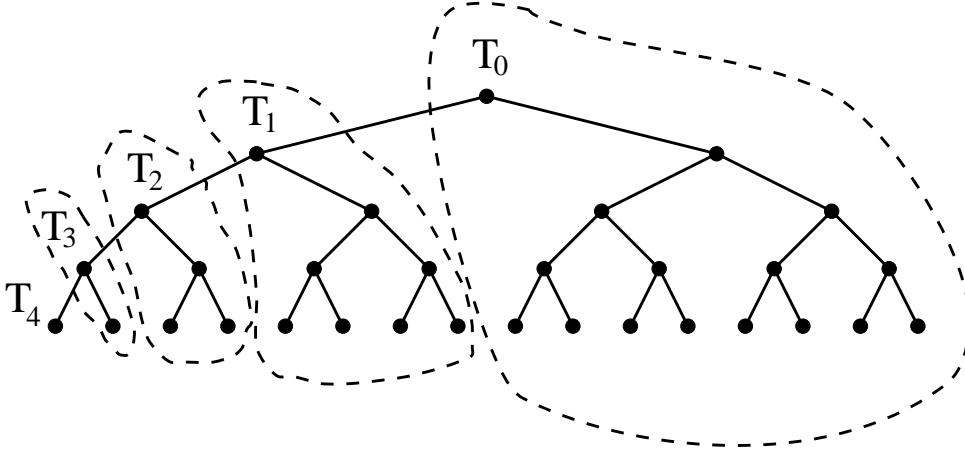,height=6cm}}
\caption{A tree with 4 generations
and a forward branching $z-1=2$. We 
denote by ${\cal T}_0$~...~${\cal T}_n$ the
vertices along a path from the top site
to a leave site. The sites descendant from
a given vertex ${\cal T}_m$ are grouped
together as shown on the figure
to calculate the generalized
Bethe-Peierls susceptibility
in section V.
}
\label{fig:tree}
\end{figure}

In order to characterize the behavior of the FIM on a hyperlattice,
we begin by considering the ordering of its central spin, 
that residing on the site deepest
in the lattice (e.g. the site ${\cal T}_0$ on
Fig.~\ref{fig:tree}).  As a point of reference, we review
the special case of the Cayley tree where the central spin
is known to undergo a mean-field Bethe-Peierls 
transition.\cite{Bethe35,Peierls36,Domb60,Baxter82,Thorpe82}.
In response to an applied uniform field on the entire tree,
its central spin develops 
the local
susceptibility $\chi^{\rm tree}_0$
\begin{equation}
\label{eq:chi-BP}
T \chi^{\rm tree}_0=\sum_l \exp{\left(
- {d_{l} \over \xi_T} \right)}
= \sum_{n=0}^{+ \infty} (z-1)^n
\exp{\left(- \frac{n}{\xi_T}\right)}
.
\end{equation}
Here we use the subscript ``0'' in
$\chi^{\rm tree}_0$ to emphasize that we are {\sl only}
considering the behavior of the central spin.
In Eq.~\ref{eq:chi-BP}, $d_l$ is the
distance between the central site ${\cal T}_0$
and the site $l$;
$\xi_T=-1/\ln{(\tanh{(\beta J)})}$
is the correlation length set by
the exponential decay of the spin correlations
$\langle \sigma_k \sigma_l \rangle
\sim \exp{(- d_{k,l} / \xi_T)}$,
identical to that 
of the Ising chain because of the
absence of loops.
The central spin susceptibility
$\chi^{\rm tree}_0$
diverges if $T<T_{\rm BP}$, with
$(z-1) \tanh{(\beta_{\rm BP} J)}=1$.
We emphasize that
this transition results from
the metric of the embedding curved space,
which leads to the 
to the prefactor
$(z-1)^n$ in Eq.~\ref{eq:chi-BP};
this is the number of sites at generation $n$ from
the central spin.
Thus, thanks to this prefactor,
exponentially decaying
correlations are sufficient to generate
mean-field ordering of the central spin.
As an aside, we note that here we have only considered
$\chi^{\rm tree}_0$, but this
calculation can be generalized to characterize the full
thermodynamic behavior associated with this spin.\cite{Baxter82,Thompson82}

For the FIM on the general hyperlattice with loops,
the expression for the local susceptibility 
is also dominated by the long length-scale contribution
of the spin-spin correlations, leading to a diverging
susceptibility with a finite correlation length.\cite{Rietman92}
Eq.~\ref{eq:chi-BP} may 
be adapted to evaluate the central-spin
susceptibility using a continuous-space description
\begin{equation}
\label{eq:chi-lob}
T \chi^{\rm lob}(0) = \int_{|z| < R} g(z)
\exp{\left(-\frac{d(z,0)}{\xi_T}\right)} d^2z
,
\end{equation}
where $g(z)$ is the metric in Eq.~\ref{eq:metrique}, and
$d(z,0)$ is the distance between
the origin and the site at 
$z = \rho \exp{(i \theta)}$ (see Eq.~\ref{eq:dist}).
We implicitly identified the distance
on the graph and the ``hyperbolic'' distance
given by the metrics. In fact, for a given distance on
the graph, there is a Gaussian distribution
of hyperbolic distances, which has been studied in
details for a Cayley tree model by
Comtet, Nechaev, and Voituriez~\cite{Comtet00}.
We claim that the physics of the FIM is not sensitive
to the existing difference between the two distances.
The resulting susceptibility
$$
T \chi^{\rm lob}(0) = 2 \pi \int_0^R
\left( \frac{1+\rho}{1-\rho} \right)^{-1/(2 \xi_T)}
\frac{\rho}{(1 - \rho^2)^2} d\rho
\sim \frac{1}{T - T_{BP}}
$$
diverges when
$\xi_T = 1/2$, with the mean-field exponent
$\gamma = 1$.
The correlation-length exponent can
be obtained by matching the Euclidean
and hyperbolic
spin correlations on short length-scales
at the Bethe-Peierls transition.
The Euclidean behavior is
$
\langle \sigma_x \sigma_y \rangle
\sim 1/|{\bf x}-{\bf y}|^{d-2+\eta}
$
with $d=2$, while the spin correlations
decay exponentially on the hyperbolic plane;
this argument leads to $\eta=0$.

We have therefore identified two critical exponents, $\gamma$ and
$\eta$,
to
have mean-field values. Scaling relations
imply that all exponents of the hyperlattice Bethe-Peierls
transition are then mean-field as well.
We note that there is another way to obtain this result
as well. On $d$-dimensional Euclidean lattices,
The upper critical dimension of spin models is related to 
the probability of intersecting random walks.\cite{Itsykson89}
the upper critical dimension of
spin models on a $d$-dimensional Euclidean lattice
is related to intersection properties of random
walks~\cite{Itsykson89}. On hyperlattices (including the special
case of trees),
the return probability of a random walk
is vanishingly small\cite{Monthus96,Cassi89,Giacometti95,Helfand83},
indicating that FIM 
hyperlattice models are above their upper
critical dimension, and thus
mean-field transitions are expected.

%
%
%

\section{Generalized Bethe-Peierls Transitions}
\label{sec:BP-general}

We would like to consider a possible Bethe-Peierls
transition for an arbitrary spin of the FIM on
a general hyperlattice.  Again, we begin by
studying the special case of loopless Cayley trees.
Here we consider spin ordering on an arbitrary site ${\cal T}_m$ at generation
$m=0$, ..., $n$ (see Fig.~\ref{fig:tree}). 
We group the sites, indicated in Fig.~\ref{fig:tree}, to expand the
susceptibility in powers of $x = \tanh{(\beta J)}$
(for details see Appendix~\ref{app:dist}) 
to obtain
\begin{eqnarray}
\label{eq:chim}
T \chi^{\rm tree}_m &=& \frac{1 - [(z-1)x]^{n-m+1}}
{1 - (z-1) x}
+ x \frac{1 - x^m}{1-x}\\
&+& \frac{ (z-2) x^2 }{1 - (z-1)x }
\left[ \frac{1 - x^m}{1-x}
-  [(z-1)x]^{n-m+1}
\frac{ 1 - [(z-1)x^2]^m}{1 - (z-1)x^2} \right]
\nonumber
.
\end{eqnarray}
In the limit $n,m \rightarrow + \infty$
and $n-m$ constant (boundary behavior),
$\chi^{\rm tree}_m(T)$ is smooth
at the Bethe-Peierls transition temperature
($(z-1) \tanh{(\beta_{BP} J)}=1$),
but diverges at a lower temperature $T'$
corresponding to $(z-1) \tanh^2{(\beta' J)}=1$.
As shown explicitly in Appendix~\ref{app:dist}, the
existence of this
transition results from a combination of
the local metric (i.e. the number of
sites a distance from the reference one)
and exponential spin correlations.
In the limit $n \rightarrow + \infty$
and $m$ fixed (bulk behavior),
$\chi^{\rm tree}_m$
in Eq.~\ref{eq:chim} diverges
at the Bethe-Peierls temperature $T_{\rm BP}$ with
mean-field behavior. If $n \rightarrow + \infty$
and $m = \lambda n$, the susceptibility
in Eq.~\ref{eq:chim} shows a divergence
at the Bethe-Peierls temperature.
The temperature $T'$ was already found by several
authors by studying the magnetic field dependence
of the Potts model on the Cayley tree~\cite{Muller,Turban}.
Here, we provide another interpretation
for this transition, in terms of a local
Bethe-Peierls transition controlled by the
metric properties at the boundary,
and prove that this transition is related to
a Griffiths phase.

We have therefore demonstrated the existence
two {\sl distinct} Bethe-Peierls transitions associated
with the bulk
(where $n-m \rightarrow + \infty$),
and the boundary, related to the
different metric properties associated with these
two different site species.  They also affect
percolation thresholds on these trees, as discussed
in 
Appendix~\ref{app:perco}.

\begin{figure}
\centerline{\psfig{file=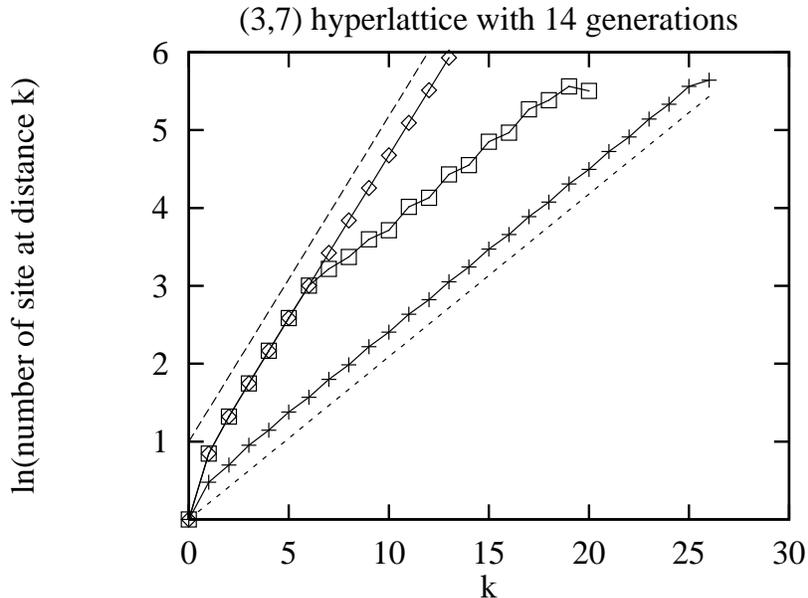,height=8.0cm}}
\caption
{Logarithm of the number of sites of the $(3,7)$ hyperlattice
with fourteen generations as a function of distance $k$ from a
given site located
(i) on the center of the hyperlattice (diamonds) where
the calculated exponential growth 
is indicated by a dashed line
(ii) on the boundary (crosses) with calculated 
exponential growth (short dashed
line)
that is the square-root of the calculated bulk growth (dashed line)
(iii) on an intermediate (7) generation (boxes), where
a crossover is observed between bulk and boundary behavior
with increasing distance scales.}
\label{fig:dist-hyper}
\end{figure}

Now that we have understood the bulk and boundary
transitions of the tree FIM  in terms of metric properties,
we continue to ask a similar
question about their more general hyperlattice analogues.
We stress that
on these manifolds of negative curvature,
site-scaling is different for reference
sites in the bulk and on the boundary.
As displayed in Eq.~\ref{eq:L-bulk}, the number of sites at a distance
$d$ from a given bulk reference one
scales like ${\cal L}_{\rm bulk} \sim (\pi/2) \exp{(2 d)}$ 
in the continuous space model.
In Appendix~\ref{app:hyp-dist} we show
the analogous quantity associated with
a boundary reference site scales like
${\cal L} \sim \exp{(d)}$.
This result is compatible with our numerical
determination of the site-scaling at a given
distance, as displayed in 
Fig.~\ref{fig:dist-hyper}.

The expression for the local susceptibility involves
a correlation length, which is not necessarily uniform for
the full lattice.
We assume a uniform, isotropic
correlation length as an Ansatz and
then
deduce the existence of bulk and boundary transitions
for the general hyperlattice set by
$\xi_T=1/2$
(see section~\ref{sec:BP})
and by $\xi_T=1$.
The existence of two different transitions
does not rely on the initial {\sl Ansatz} because
the correlation length at the boundary
can only be reduced compared to that of the bulk.
Therefore,
the initial {\sl Ansatz} of uniform
correlations indicates
the existence of a boundary transition
at a temperature lower than the bulk transition.

\begin{figure}
\centerline{\psfig{file=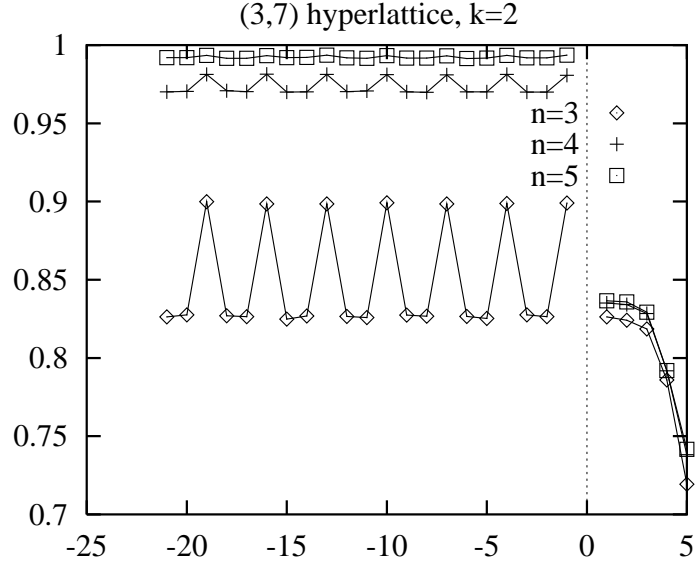,height=7.5cm}}
\caption{Spin-spin correlations at $T=1$ for
the central (negative x-axis) and a boundary spin (positive x-axis)
at a distance $k=2$ where $x$ is an arbitrary spin index.
}
\label{fig:corre-bulk}
\end{figure}

\begin{figure}
\centerline{\psfig{file=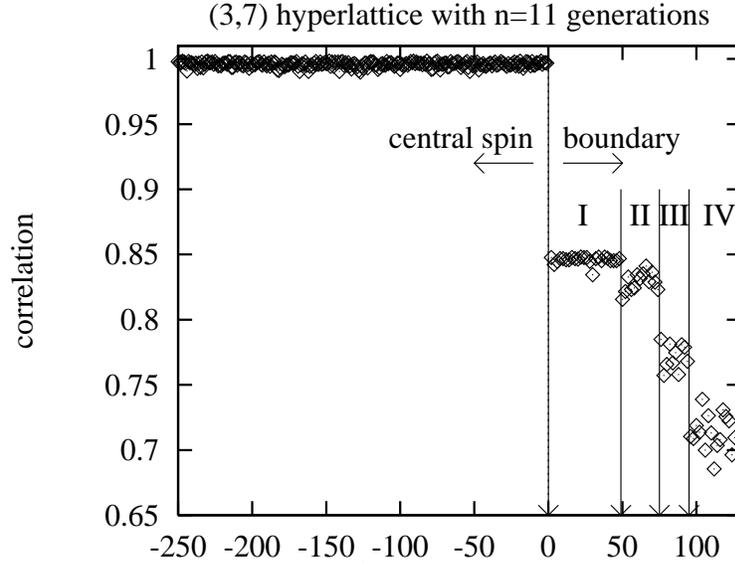,height=7.5cm}}
\caption{Spin-spin correlations
computed for the {\sl central} and a boundary
spin
at a distance $k=7$.
}
\label{fig:corre-bound}
\end{figure}

We support the previous analysis, assuming a uniform spin correlation
length, with numerical determinations of the spin-spin correlations.
On the Cayley tree, 
the correlation
of two spins $\sigma$ and $\sigma'$ at a distance $d$ 
is $[\tanh{(\beta d)}]^d$, irrespective of the site locations
of the two reference spins.
By contrast, as discussed below,
the situation is different for the general hyperlattice case
where the correlations are reduced at the boundary.
We distinguish between spin correlations in the
bulk and at the boundary:

{\sl Bulk}: The symmetry group of
the $(3,7)$ hyperlattice with a finite number 
of layers $n$
is generated by one rotation of
angle $2 \pi /7$ around the central spin, 
and one inversion with respect to an axis going through
the central spin. The infinite hyperlattice has a
huge symmetry group: the lattice is left invariant
under the aforementioned set of symmetries around
{\sl any} lattice site, a fact that is no longer true for
a finite system.
However the properties of the finite-size lattice
reflect the huge symmetry of the infinite lattice.
To be more precise, let us consider a FIM on a hyperlattice
with $n$ layers and $N_k$  correlations
$\langle \sigma_0 \sigma_k \rangle$ between
a given spin $\sigma_0$, and any of the $N_k$
spins at a distance $k$ from $\sigma_0$.
Let us first take
$\sigma_0$ 
to be the central spin.
Strictly speaking,
the number of such different correlations
is $N_k/7$ because of the $2 \pi /7$ rotation
symmetry. However we find strong evidence that,
for $k$ fixed and $n$ large, {\sl all}
the $\langle \sigma_0 \sigma_k \rangle$
correlations nearly 
coincide.
On the lower, negative x-axis of
Fig.~\ref{fig:corre-bulk}, we
have chosen $k=2$ and increased $n$; the effect
of the site alternation with coordination
3 and 4 is clearly
visible (see the boundary on Fig.~\ref{fig:37}).
The correlation inhomogeneities 
are reduced with increased system size (cf. the negative $x$-axis part of
Figs.~\ref{fig:corre-bulk}
and~\ref{fig:corre-bound}).

{\sl Boundary behavior}:
Now we consider the reference spin $\sigma_0$ to be
located at the hyperlattice boundary
and the spins $\sigma_k$ at a distance $k$
from $\sigma_0$.
The dispersion is similar to that already
described for 
bulk case, though       
it persists in the
large-$n$ limit (see
the positive $x$-axis part on
Fig.~\ref{fig:corre-bound}).
The sites with smaller coordination are less correlated,
as expected on physical grounds. 
Therefore the correlations
are weakened at the boundary of the hyperlattice.

\section{Magnetization Distributions at Low Temperatures}
\label{sec:PM}

\begin{figure}
\centerline{\psfig{file=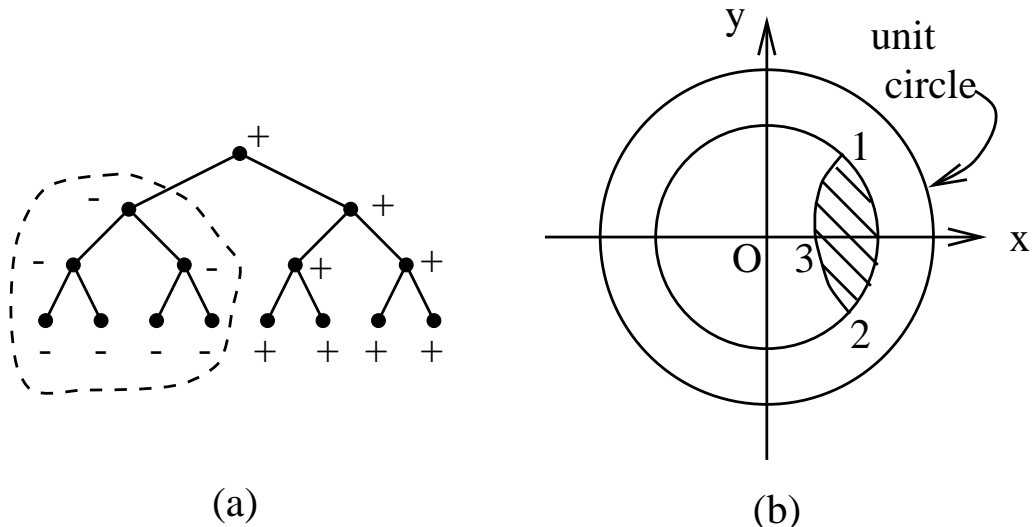,height=7cm}}
\caption{Magnetic domains (a) on the tree, and (b) on
the hyperlattice. The geodesic
1-3-2 is perpendicular
to the border of the hyperlattice $R \exp{(i \phi)}$.
Point 1 is $R \exp{(i \theta)}$. Point 2 is
$R \exp{(-i \theta)}$
}
\label{fig:domain}
\end{figure}

We now demonstrate slow dynamics at low temperatures
in this periodic hyperlattice FIM.
Towards this goal,we track the energy and magnetization distributions,
comparing results obtained using two sampling methods.
The first is a standard heat bath single spin-flip algorithm
that probes configuration space where two configurations are
considered neighbors if they differ by exactly one spin.
The second one is a cluster algorithm that can go 
from any configuration to any other in just one step,
and is applicable here because of the absence of frustration.
We use these
two algorithms to probe the structure
of the configuration space, analogous
to similar studies performed previously
on Cayley trees.\cite{Melin96,Melin96a}
In both cases,
the valleys correspond to magnetic domains in real space
(see Fig.~\ref{fig:domain}). On the hyperlattice, the
energy of a magnetic domain scales like the logarithm
of its area (see Appendix~\ref{app:vol}) while it
does not scale with area on the tree.

\begin{figure}
\centerline{\psfig{file=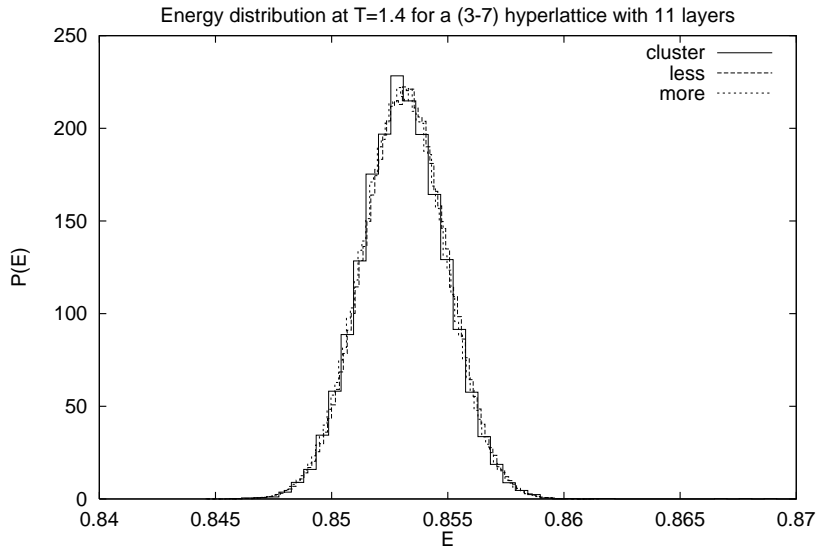,height=7.5cm}}
\caption{Energy distribution computed at $T=1.4$
using single-spin flip cluster algorithms
where ``more'' and ``less'' curves correspond to 47800 and
300000 MCS/spins respectively; the ``cluster'' curve
was obtained with 10200 updates of the cluster algorithm.
}
\label{fig:dise}
\end{figure}

In Figure~\ref{fig:dise}, we present three energy distributions.
One was acquired with a cluster algorithm, while the other two
were obtained with the single-spin flip algorithm and $3\times 10^5$
and $4.78 \times 10^5$ MCS-spins. The three simulations were performed
at a temperature $T=1.4$, below the estimated Bethe-Peierls
transition (of order of the lattice coordination).
These three energy distributions
are close to 
the same Gaussian distribution,
centered at the mean value of the internal
energy, with a width proportional to the specific heat
times the square of the temperature.
>From this figure it is clear that
there are no long time-scales present as
far as the energy is concerned.

\begin{figure}
\centerline{\psfig{file=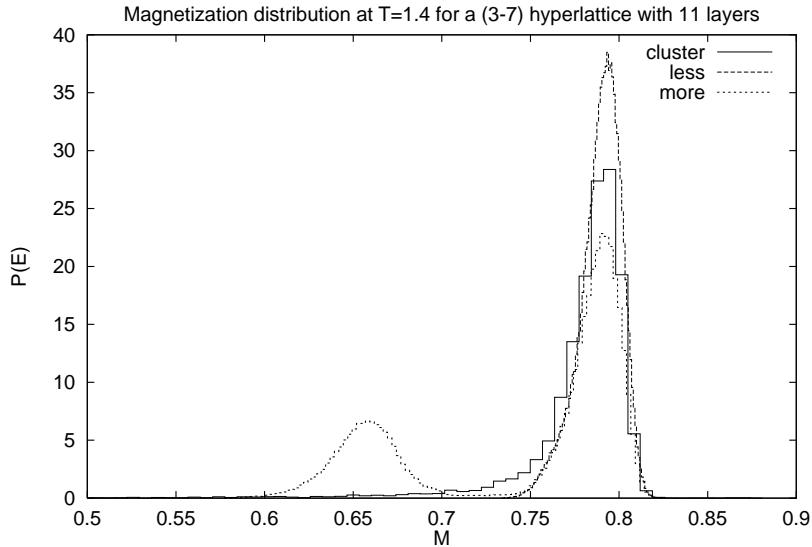,height=7.5cm}}
\caption{Magnetization distribution computed at $T=1.4$
using single-spin flip and cluster algorithms
where ``more'' and ``less'' curves correspond to 47800 and
300000 MCS/spins respectively; the ``cluster'' curve
was obtained with 10200 updates of the cluster algorithm.
}
\label{fig:dism}
\end{figure}

However, the behavior of the magnetization is rather different.
In Fig.~\ref{fig:dism} we present three magnetization distributions,
determined during the same runs as
the internal energy distributions.
The three magnetization distributions
are {\sl distinctive}. The two distributions obtained using
the single-spin flip algorithm are {\sl bimodal}
whereas the distribution
obtained with a cluster algorithm is {\sl unimodal}. For
other runs, with less statistics, the magnetization distribution
with the single spin flip algorithm
was found to be multi-modal, with three or more maxima.
This suggests the existence of energy barriers
which trap the system with single-spin dynamics.
These simulations provide evidence that the low temperature
dynamics are slow.
The glass crossover temperature
corresponds to the correlation length being comparable
to the system size. Also the magnetization distributions
are broad at low temperature (see Fig.~\ref{fig:dism})
for the hyperlattice, similar to that found for the
Cayley tree;\cite{Melin96} this is an indication
of relaxation on many time-scales.
This is because of the stability 
of a Griffiths phase on the entire lattice, at
a temperature set by the boundary Bethe-Peierls transition
(see Appendix~\ref{app:Griffiths}).

\section{Discussion}
\label{sec:conclusion}

In summary, we have studied the nearest-neighbor
ferromagnetic Ising model
on tilings embedded in hyperbolic surfaces with
negative curvature, often beginning with the special
case of Cayley trees.
We have identified two mean-field transitions
in these systems, associated with ordering
of the bulk and the boundary spins as a function
of decreasing temperature.
These two transitions can be understood in
terms of the local metric properties of the
lattice, specifically by the distinct scaling of
sites with distance from the boundary and bulk spins.
We believe that these two transitions will be
characteristic of
{\sl any} short-range
ferromagnetic 
model residing on a planar surface with globally negative curvature 
independent of lattice specifics or
the nature of the spin order parameter.

In order to satisfy the criterion of
``global negative curvature'', 
at large length-scales the resident lattice should be embedded on
a surface with a negative (not necessarily constant) curvature 
instead of on a Euclidean plane.
For instance, we could imagine models
with a fluctuating metric. As a simple
example, we could consider a
recursive random lattice
in which the sites have a coordination $z$ with
a given probability $P(z)$. The sites with $z=2$ have the local
environment of an Ising chain, and the
sites with $z \ge 3$ have local branching.
The bulk Bethe-Peierls transition
is found to be $ \LL z-1 \RR
\tanh{(\beta_{\rm bulk} J)}= 1$,
whereas the boundary transition corresponds to
$\sqrt{\LL z-1 \RR} \tanh{(\beta_{\rm bound} J)}=1$.
Both transitions occur at a finite temperature
as soon as $\LL z \RR >2$, in which case
the metric is globally 
hyperbolic. 

\begin{figure}
\centerline{\psfig{file=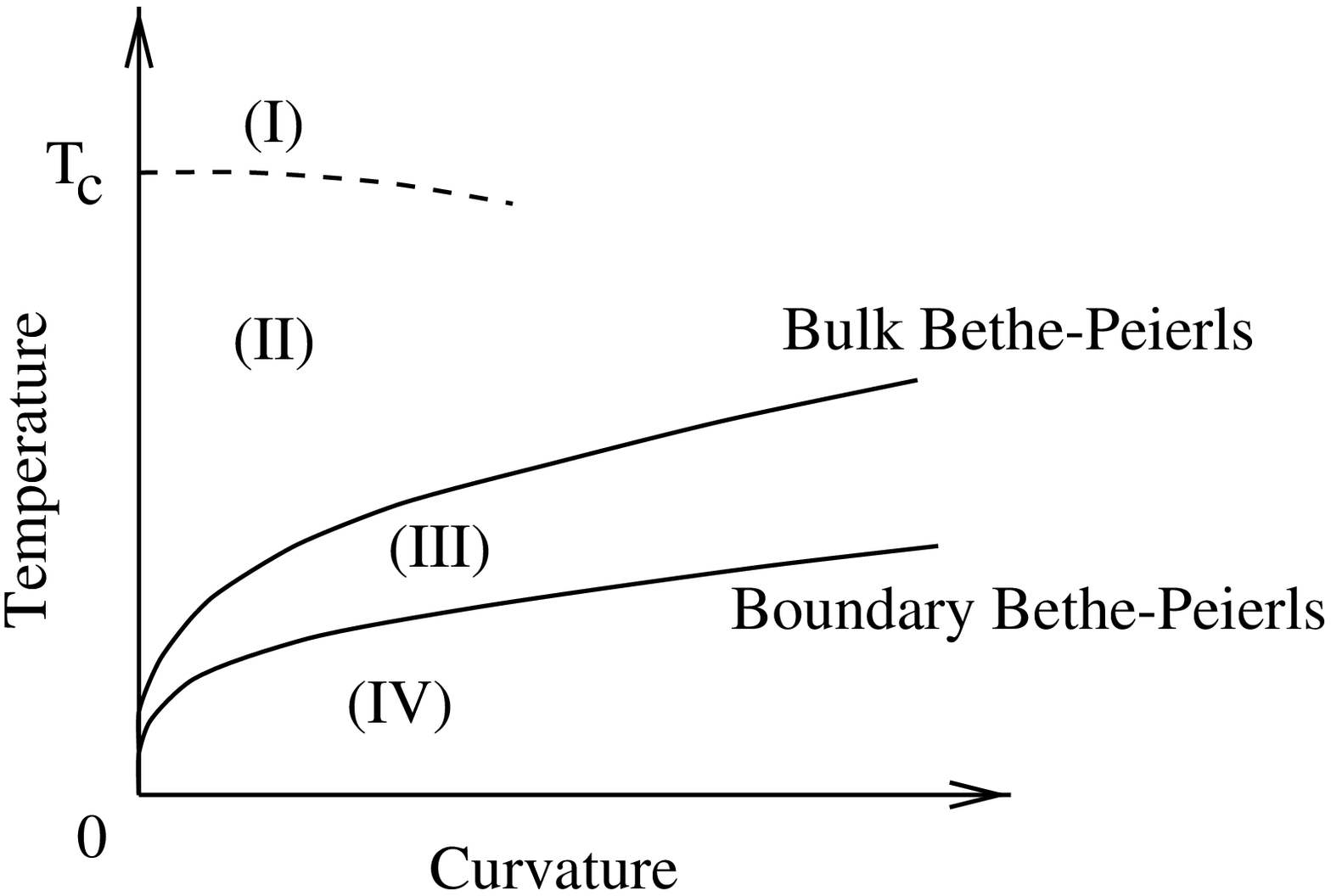,height=7cm}}
\caption{(Curvature, Temperature) phase diagram.
$T_c$ is the critical temperature of the Euclidean
model. The dashed line is a cross-over.
Region (I) is the paramagnet. Region (II)
corresponds to algebraic correlations below the
curvature radius ${\cal R}$, and exponential correlations
above ${\cal R}$. The solid lines are the bulk and
boundary Bethe-Peierls transitions, determined
as $\xi_T = {\cal R}$ and $\xi_T = 2 {\cal R}$
respectively. Note that $\xi_T$ is related to the
exponential decay of the correlations above the
curvature radius, and {\sl not} to the behavior
of the correlations below the curvature radius.
Region (IV) corresponds to the Griffiths phase
of the entire lattice.
}
\label{fig:Curvature-T}
\end{figure}

One can also question how the hyperbolic/Euclidean
transition
would operate in a continuous model in the limit of a
large curvature radius ${\cal R} \gg 1$.
The bulk and boundary Bethe-Peierls transitions
would be of order $1 / {\cal R}$ while Euclidean
physics would develop below the length scale ${\cal R}$.
Strictly speaking, the Euclidean phase transition
occurs only when ${\cal R} = + \infty$.
However, with a finite but large ${\cal R}$,
we expect a pronounced maximum in the susceptibility
and specific heat. The expected behavior is
summarized on Fig.~\ref{fig:Curvature-T}.
Again we emphasize that we expect this phase behavior
for ferromagnetic models
residing on hyperboloids
independent of the details of the spin order parameter.
For Cayley trees, it has been shown that the nearest-neighbor
$xy$ model
displays similar physics (i.e. mean-field transition) as
its Ising counterpart.\cite{Melin96}  
For a general
hyperlattice, we expect two mean-field transitions
for short-range $xy$ models since there exist loops and
two species of
spin (boundary and bulk), as in the Ising case.
As was done in this Paper, these transitions would be obtained
by comparing the exponential decay
of the correlations to the growth in the
number of sites with distance; this would be interesting to
verify..

The physics of the low-temperature behavior of the hyperlattice
FIM is determined by the formation of droplet-like excitations
nucleating from the boundary.  These rare fluctuations do
not affect the specific heat (Section III), but do lead
to a broad magnetization distribution reflecting a wide spectrum
of relaxation time-scales.  Such Griffiths phases are usually
associated with dilute ferromagnets with spatial inhomogeneity.
In the periodic models we have studied here there is no intrinsic
disorder; however there are two distinct site species, a feature
that contributes to favoring droplet formation. Indeed here we
report a diverging susceptibility per site at the boundary
transition(see
Appendix~\ref{app:Griffiths}), 
a characteristic shared with random ferromagnets
with correlated disorder\cite{Pujol00}.  Similarly divergences
are reported for quantum disordered magnets, where
the randomness is correlated in the time-dimension.
To our knowledge, this is the first time a Griffiths
phase have been identified in a periodic magnetic system
with loops;
furthermore the identification of slow dynamics in a short-range
periodic system at low temperatures is very encouraging.

Though motivated by experiment, our chosen
lattices of study are somewhat removed from the
structures commonly observed in Nature.
However perhaps we can close by attempting
to reconnect with experiment.  In particular,
we have identified Griffiths phases in ferromagnetic
models residing on hyperlattices.  Though this
phenomenon has been discussed extensively in
the theoretical literature, it
has not yet been conclusively identified in the laboratory.\cite{Mydosh00}
For example, there have been claims in random-field materials,\cite{Binek94}
but they remain controversial due to plausible alternative
interpretations of the data.\cite{Dotsenko94}
Perhaps Josephson junction arrays fabricated in a hyperlattice
topology would provide a promising setting for the observation
of this well-discussed phenomenon. In this artificial network,
a junction is
located at each link of the hyperlattice; the short superconducting
wires  
have a phase $\phi_i$ with a Hamiltonian
$$
H = E_J \sum_{\rm \langle i,j \rangle} [1 - \cos{
(\phi_i - \phi_j)}]
.
$$
where the ground-state configurations corresponds
to a uniform $\phi_i=0$ and $E_J$ is an energy-scale associated
with a junction. 
If this array were placed in a time-dependent transverse magnetic field,
the low-frequency part of the resulting ac 
susceptibility\cite{Chandra97} should provide 
a useful probe for the slow dynamics associated with the Griffiths
phase.

In order for our ideas to be applicable to the Josephson hyperlattice,
it
should not display a vortex binding-unbinding transition.
Assuming that the nature of the spin order parameter is not
important, we can apply some of our previous expressions to
check this.
The phase at a distance $d$ away from
a given bulk vortex scales like
$\phi_{\rm bulk}(l) \sim 4 \exp{(- 2 l)}$, where we have
used the
expression Eq.~\ref{eq:L-bulk} for the number of sites at
a given distance $l$. The energy of
a junction a distance $d$ away 
from a bulk node
$\frac{1}{2} E_J [\phi(d)]^2$, leading to the
total vortex energy $E_{\rm bulk} = 2 \pi E_J$.
It is remarkable that this quantity is {\sl finite},
since it diverges logarithmically with distance
on the square lattice.
Similarly, the energy of a vortex at the boundary
is found to be $E_{\rm bound} = 2 \pi^2 E_J$,
again a finite quantity. As a consequence,
on the hyperlattice,
there is no Kosterlitz-Thouless transition associated
with the unbinding of vortex-antivortex pairs.
We note that an identical conclusion was reached
for two-dimensional Coulomb systems residing
on a surface of constant negative curvature;\cite{Jancovici98}
this agreement gives us further confidence in our conjecture
about the universal nature of the phase behavior that we
have identified.

\section*{Acknowledgments}
One of us (R.M.) acknowledges a fruitful discussion
with A. Comtet, S. Nechaev,
and R. Voituriez.

\newpage

\appendix
\section{Number of sites at a given distance on the tree}
\label{app:dist}
We consider a site at generation $m$ on a tree
and calculate the number of sites
$N_m(l)$ at a given 
distance $l$ from the site at generation $m$.
This is given
by the prefactor of the term $x^l$ in
the expansion of the
susceptibility in powers of $x = \tanh{(\beta J)}$:
\begin{eqnarray}
\label{eq:chi-tree-exp}
T \chi_{\rm BP}^{\rm tree}(m) 
&=& \frac{1-[(z-1)x]^{n-m+1}}{1-(z-1)x}\\
&+&
\sum_{k=0}^{m-1} x^{m-k} \left[ 
1+ (z-2) x + (z-2)(z-1) x^2 + ...
+ (z-2)(z-1)^{n-k-1} x^{n-k} \right]
.\nonumber
\end{eqnarray}
Eq.~\ref{eq:chim} has been obtained by summing
the series in Eq.~\ref{eq:chi-tree-exp}.
First the sum $\sum_{l=0}^{n-m} [(z-1)z]^l$
gives rise to $(z-1)^l$ sites at a distance
$0 \le l \le n-m$. Next, the sum
$\sum_{k=0}^{m-1} x^{m-k}$ results
in one site at a distance $1 \le l \le m$.
The remaining terms give rise to 
$(z-2) (z-1)^{l-m+k-1}$ sites at a
distance $l$, with $0 \le k \le m-1$
and $m-k+1 \le l \le n + m- 2 k$.
We therefore need to distinguish between
two cases: case (I) $m \ge \left[
(n-1)/2 \right]$; and case (II)
$m \le \left[ (n-1)/2 \right]$,
with $\left[ ... \right]$
the integer part.

\begin{itemize}
\item{Case (I): $m \ge \left[(n-1)/2 \right]$}.
We should compare $l$ with $n-m$ and
$m+1$. We find three cases:
\begin{itemize}
\item{(i) $l \le n-m$.} The number of sites
at distance $l$ is $(z-1)^l-1$.

\item{(ii) $n-m \le l \le m+1$.} The number of 
sites at distance $l$ is 
$(z-1)^{\left[ (n-m+l)/2 \right]} -1$.

\item{(iii) $m+1 \le l$.} The number of 
sites at distance $l$ is 
$(z-1)^{\left[ (n-m+l)/2 \right]}
- (z-1)^{l-m-1}$.

\end{itemize}

\item {Case (II): $m \le \left[ (n-1)/2 \right]$}. We compare
again $l$ with $n-m$ and $m+1$:
\begin{itemize}

\item{(i) $l \le m+1$.} The number of 
sites at distance $l$ is 
$(z-1)^l - 1$.

\item{(ii) $m+1 \le l \le n-m$.} The number of 
sites at distance $l$ is 
$(z-1)^l - (z-1)^{l-m-1}$.

\item{(iii) $n-m \le l$.} The number of 
sites at distance $l$ is 
$(z-1)^{\left[ (n-m+l)/2 \right]}
- (z-1)^{l-m-1}$.
\end{itemize}

\end{itemize}
Therefore at large distance the number of
sites at a distance $l$ from a given site
at generation $m$ scales like
$N_m(l) \sim (z-1)^{l/2}$
if $m \ge \left[(n-1)/2 \right]$. This
scaling should be compared with the
exponential decay of the correlations
$\langle \sigma_i \sigma_j \rangle
\sim \exp{(- d_{i,j} / \xi_T)}$
and leads to a boundary
transition at a temperature
$T'$ given by
$(z-1) \tanh^2{(\beta' J)}=1$. This
transition occurs whenever $n-m$ is
finite in the limit $n \rightarrow
+ \infty$.

\section{Percolation on the tree: generalized Bethe-Peierls
limit}
\label{app:perco}
We would like
to illustrate the different behavior of the
generalized Bethe-Peierls transitions on the
example on bond percolation where the 
same phenomenon occurs as in the case
of ferromagnetism.
We denote by $P^{\rm BP}_n(M)$ the
probability to find $M$ sites
(a mass $M$)
in the cluster containing the top
site of a tree with $n$ generations.
We have
$$
P_{n+1}(M) = \sum_{\theta_1} ... \sum_{\theta_{z-1}}
\sum_{M_1} ... \sum_{M_{z-1}}
\prod_{i=1}^{z-1} P_n(M_i)
\prod_{i=1}^{z-1} p(\theta_i)
\delta \left( M - \sum_{i=1}^{z-1} \theta_i M_i -1 \right)
,
$$
with $\theta_i=1$ with a probability $p(1)=\mu$
and $\theta_i=0$ with a probability $p(0)=1-\mu$.
The average number of sites in the cluster
containing the top spin of a tree with $n$
generations is iterated as
$\tilde{M}_{m+1} = (z-1) \mu \tilde{M}_n + 1$.
The fixed point
value is $\tilde{M}^{*} = 1/(1 - (z-1) \mu)$ and
diverges at the bulk percolation threshold
$\mu_P^{\rm 0}=1/(z-1)$.

We now consider the number of sites
$M({\cal T}_m)$
in a cluster
containing the point ${\cal T}_m$ 
in the presence of $n$ generations
(see 
Fig.~\ref{fig:tree}). We find
\begin{eqnarray}
M({\cal T}_m) &=&
\frac{
[(z-1) \mu]^{n-m+1} -1}
{(z-1) \mu -1}\\
&+& \frac{ (z-2) \mu^2} {(z-1) \mu -1}
\left(
[(z-1) \mu]^{n-m+1}
\frac{ [(z-1) \mu^2]^m -1}
{ (z-1) \mu^2 -1}
- \frac{ 1 - \mu^m}{1-\mu}
\right)
\nonumber
,
\end{eqnarray}
of the same form as Eq.~\ref{eq:chim}.
If $n,m \rightarrow + \infty$ with
$n-m$ constant, the cluster size
containing the vertex ${\cal T}_m$
does not diverge at the bulk
percolation threshold $\mu_P^{ 0}$
while it diverges at a larger percolation
threshold $\mu_P'=1/\sqrt{z-1}$. If
$n \rightarrow + \infty$ and $n-m$
growing faster than $n$,
we find a percolation
transition at the bulk
percolation threshold $\mu_P^{ 0}$.
Therefore, similarly to ferromagnetism, percolation
shows a different behavior depending on how
the Bethe-Peierls limit is taken. The
boundary percolation threshold $\mu_P'$
is larger than the bulk percolation
threshold $\mu_P^0$.

\section{Number of sites at a given distance from
a boundary site on the hyperlattice}
\label{app:hyp-dist}

We consider a finite size disc $|z|<R$
with the hyperbolic metrics Eq.~\ref{eq:metrique},
and calculate the length of the set of points
at a distance $d$ from the boundary site
at coordinate $R$. This set of points
is represented by the circle $x_0 + R_0
\exp{(i \phi)}$ (see Eq.~\ref{eq:x0-R0}),
which intersects
the boundary $|z|=R$ at the points
$z^{\pm} = x_0 + i R_0
\exp{(\pm i \phi_0)}$, with
$$
\cos{\phi_0} = 
\frac{ \tanh{d}}{2 R (1 - \tanh^2{d})}
\left( 1 + (-3 + \tanh^2{d}) R^2
+  (\tanh^2{d}) R^4 \right)
.
$$
A straightforward calculation leads to the length
of the arc $z^+$--$z^-$:
$$
{\cal L}_{\rm bound} =
\int  \sqrt{ g(z) }  R_0 d \phi
=
\frac {4 \tanh{d}}
{1 - \tanh^2{d}}
\tan^{-1}{ \left[
\sqrt{ \frac{ 2 R - (1 + R^2) \tanh{d}}
{2 R + (1 + R^2) \tanh{d}}}
\right]}
.
$$
In the relevant regime $1 \ll d \ll \tanh^{-1}{R}$
with $R \simeq 1$,
we find ${\cal L}_{\rm bound} \sim \exp{(d)}$. This boundary
behavior should be contrasted with the bulk
behavior ${\cal L}_{\rm bulk} \sim
(\pi / 2) \exp{(2 d)}$ (see Eq.~\ref{eq:L-bulk}).

\section{The length of magnetic domains}
\label{app:vol}
\begin{figure}
\centerline{\psfig{file=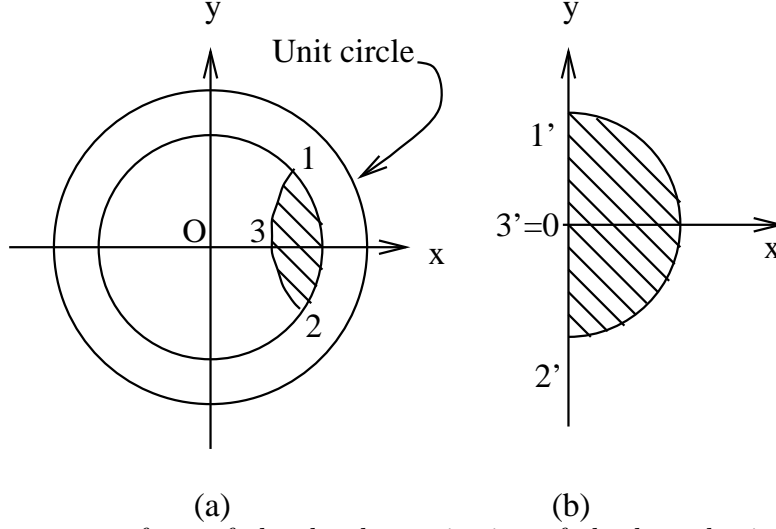,height=7cm}}
\caption{Isometry transform of the droplet excitation
of the hyperlattice. (a) the original droplet,
and (b) the transformed droplet. The isometry is chosen such as
the segment 1'-3'-2' is on the $y$ axis with
3' at the origin. Point 1 is $R \exp{(i \theta)}$,
point 2 is $R \exp{(-i \theta)}$.
Point 1' is $i \rho$,
point 2' is $-i \rho$.
}
\label{fig:isometry}
\end{figure}
We show that the length of the droplet excitations
shown on Fig.~\ref{fig:domain}(b) is proportional
to the logarithm of its area. The difference with
the calculation in Appendix~\ref{app:hyp-dist} is
that we consider here the geodesics 1--3--2 corresponding
to the physical situation where the
energy of the domain wall is minimized.

To
calculate the shaded area on Figs.~\ref{fig:domain}(b)
and~\ref{fig:isometry}(a), and use the transformation
$
z \rightarrow z'=(a z + b)/(b z + a)
$
with $a$ and $b$ real numbers, to
map the
domain on Figs~\ref{fig:isometry}(a)
and~\ref{fig:domain}(b) 
into the domain on Fig.~\ref{fig:isometry}(b).
Imposing point 1 to be transformed into 1'
at coordinate $i \rho$ leads to
$$
\alpha \equiv {b \over a} 
= - \frac{ R \cos{\theta} (1 + \rho^2)}
{1 + \rho^2 R^2 + 2 \rho R \sin{\theta}}
,
$$
with
$$
\rho = \frac{1}{2 R \sin{\theta}}
\left( -(1-R^2) + \left(
(1-R^2)^2 + 4 R^2 \sin^2{\theta} \right)^{1/2} \right)
.
$$
The area element is found to be
$$
d {\cal A} = g(r') d r' d \theta' =
\frac{r}{2(1 - r^2)}    
\frac{ (1+\alpha^2)r 
+ (r^2+1) \alpha \cos{\theta}}
{1 + \alpha^2 r^2 + 2 \alpha r \cos{\theta}} d \theta
.
$$
A straightforward 
integration over $\theta$ leads to
$$
{\cal A} = \frac{\theta}{2} \left( \frac{1+R^2}{1-R^2}
\right) + 
\frac{\alpha^2 R^2-1}{|1 - \alpha R|}
\tan^{-1}{\left( \left|  \frac{1 - \alpha R}
{1 + \alpha R} \right| \tan{ \left( {\theta \over 2}
\right)} \right)}
.
$$
When $R \simeq 1$, we have
$
{\cal A} \sim {\theta}/{[2(1-R)]}       
$.
The length of the line 1--3--2 is
$
{\cal L} = \frac{1}{2} \ln{
[({1 + \rho})/({1 - \rho})]}
 \sim - {1 \over 2} \ln{(1-R)}
$.
The scaling between the
area and length of a magnetic domain
in the limit $R \rightarrow 1$ is
finally found to be
$
{\cal A} \sim \frac{ \theta}{2} \exp{(2 {\cal L})}
$: the area is exponentially large in the
boundary length.

\section{Total susceptibility: Griffiths transition}
\label{app:Griffiths}

We derive the total susceptibility
of the tree and hyperlattice models.
The susceptibility of both models diverges below the
Griffiths temperature set by the local transition
temperature of the boundary.

\subsection{Tree model}

The susceptibility of the entire lattice
$\chi_{\rm TOT}$ is obtained as the sum
of the local susceptibilities:
$
\chi_{\rm TOT} = \sum_{m=0}^n (z-1)^m \chi_m(T)
$.
First if $T > T'$, the susceptibility per spin
behaves like $\chi_{\rm TOT} / N_n \sim
(1+x)^2 / [1 - (z-1) x^2]$, and diverges at the
temperature $T'$. Now if $T < T'$, the susceptibility
per spin behaves like $\chi_{\rm TOT} / N_n \sim
[(z-1) x^2]^n$, an infinite quantity in the
limit $n \rightarrow + \infty$.

\subsection{Hyperlattice model}

We first estimate the local susceptibility
of the hyperlattice with a finite size
$|z| < R$
at the point represented by a real
number $x$, and
we assume the correlations to be uniform
as discussed in Section V of the main body
of the text.
We note $d = d(x,R) = \tanh^{-1}{
[(R-x)/(1 - xR)]}$ the distance between the points
$x$ and $R$, and $D = \tanh^{-1}{
[(R + x) / (1 + xR)]}$ the distance between
the points $x$ and $-R$.
Based on Fig.~\ref{fig:dist-hyper} and
Appendix~\ref{app:hyp-dist}, we approximate
the growth of the number of sites ${\cal N}(l)$
at a distance $l$ away from the point $x$
as follows:
(i) $l < d$ (bulk behavior):
${\cal N}(l) \simeq (\pi/2) \exp{(2 l)}$;
(ii) $d < l < D$ (boundary behavior):
${\cal N}(l) \simeq (\pi/2) \exp{(l+d)}$.
The local susceptibility is found to be
\begin{eqnarray}
\label{eq:sus-toy-model}
T \chi(x) &\simeq & \int_0^D {\cal N}(l) \exp{(- l
 / \xi_T)} dl \nonumber \\
&=& \frac{ \pi}{2} \frac{1}{2 - 1 /\xi_T}
\left[ e^{(2 - \frac{1}{\xi_T})d} - 1 \right]
+
\frac{\pi}{2} \frac{1}{1 - 1 / \xi_T}
e^{d} \left[ e^{(1 - \frac{1}{\xi_T})D}
- e^{ (1 - \frac{1}{\xi_T})d} \right]
.
\end{eqnarray}
In the case of the central spin
$x=0$, we find a susceptibility
diverging at $\xi_T = 1/2$. In the
case of a boundary site $x=R$,
we find a susceptibility diverging
at $\xi_T = 1$, consistent with the behaviors
obtained in the main text of this paper.
The total susceptibility is obtained as
the sum of the local susceptibilities
Eq.~\ref{eq:sus-toy-model}:
\begin{equation}
\label{eq:chi-tot-lob}
\chi_{\rm TOT} = \int_0^R 
2 \pi x g(x) \chi(x) dx
.
\end{equation}
We note $R = 1- \epsilon_R$, and expand
the susceptibility Eq.~\ref{eq:chi-tot-lob}
in the parameter $\epsilon_R \ll 1$.
The number of sites $N_s$ scales like $\epsilon_R^{-1}$.
If $\xi_T<1$, the susceptibility per site
is found not to
scale with $N_s$. If $\xi_T>1$, the susceptibility
per site is found to scale
like $\epsilon_R^{-1 + 1 / \xi}$,
an infinite quantity in the limit $\epsilon_R \rightarrow
0$.

Therefore, the tree and hyperlattices models
show the same behavior: the susceptibility
per site is infinite below the Griffiths transition
temperature $T'$. The Griffiths temperature $T'$
is equal to the temperature of the boundary
Bethe-Peierls
transition. Physically, this originates because
of the finite fraction of spins at the boundary:
the boundary dominates the
behavior of the total susceptibility.

\end{document}